1

# A Compact Dynamic Antenna for Physical Layer Wireless Security

Sheng Huang, *Member, IEEE,* Jacob R. Randall, *Graduate Student Member, IEEE*, Cory Hilton, *Graduate Student Member, IEEE*, Jeffrey A. Nanzer, *Senior Member, IEEE*

*Abstract*—We propose a novel omnidirectional antenna design incorporating directional modulation for secure narrow planar information transmission. The proposed antenna features a compact size and stable omnidirectional radiation performance by employing two tightly spaced, printed meander line monopole antennas, acting as a single radiating element. To achieve a narrow information secure region, the antenna is fed by differential power excitation of two ports with real-time dynamic switching. This leads to phase pattern modulation only along the electrical polarization, resulting in a directionally confined region where information is recoverable in the E-plane, while maintaining a highly constant or static omnidirectional H-plane pattern, inducing a 360° information recoverable region. The dynamic antenna is designed and fabricated on a single layer of Rogers RO4350B which provides a miniaturized planar size of $0.36 \times 0.5 \lambda_0^2$ at 2.7 GHz. The fabricated antenna is directly fed with a 10 dB power ratio by a radio frequency (RF) switching system and evaluated for 16-QAM and 256-QAM transmission in a high signal-to-noise ratio (SNR) environment. For 16-QAM transmission, a narrow E-plane information beam (IB) of approximately 34° and omnidirectional H-plane IB are obtained, and a narrower E-plane IB is achieved around 15° for 256-QAM.

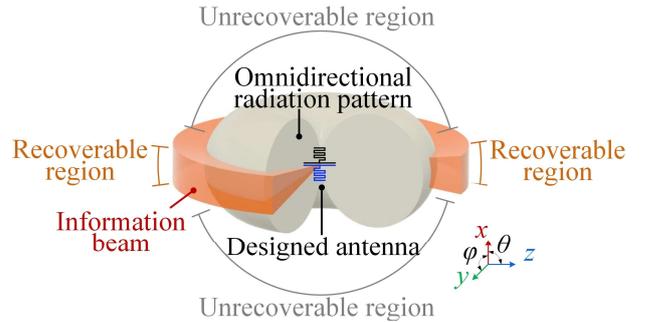

**Fig. 1.** The proposed antenna enables spatially selective information transmission, where only receivers located within a narrow sector in $\varphi = 0°$ plane can recover the modulated signal. In the $\varphi = 90°$ plane, the antenna maintains omnidirectional radiation without directional modulation, allowing information to be recovered across all angles.

## I. INTRODUCTION

IN emerging applications such as the Internet of Things (IoT), short-range wireless networks, and unmanned communication systems, traditional cryptographic methods face significant limitations due to constraints in device resources, computational complexity, and latency [1]. As a result, physical-layer security (PLS) has attracted growing interest as a complementary solution, enabling security directly at the physical level [2], [3]. One such approach is directional modulation (DM) which has emerged as a promising technique due to its spatial selectivity via modulating the amplitude and phase of the radiation pattern, encoding radiation characteristics that are only decodable within specific angular regions.

Conventional DM implementations, however, typically rely on complex phased arrays that require multiple RF chains, precise beamforming control, and bulky feed networks [4]-[9]. To overcome these limitations, some studies have investigated simplified DM architectures, such as near-field antenna modulation [10] and distributed arrays for increased efficiency [11]. The work presented in [12] explored the DM on a single element and the work in [13] investigated the DM for omnidirectional modulation with array employment. However, these works suffer from limited secure beamwidths and complicated designs. It is seen that the research for compact antenna design and simple DM capability is lacking, and will be of increasing relevance as the implementation of wireless communication devices rapidly expand. In terms of antenna miniaturization, the exploration of meander line structures is one of the most popular methods [14]–[19]. The meander line antenna consists of periodic folded conductors, leading to a lowered resonance frequency with extended electrical length compared with the straight dipole antenna [20], [21]. Many meander line based works have been reported in various applications including portable devices [22], wearable antennas [23], [24], and RF identification tags [25]–[32], as well as designs of system-in package antennas [33].

This letter demonstrates a low-cost and compact antenna design for near-omnidirectional radiation characteristics with dynamic phase modulation in the electrical polarization to enhance the planar information security, as illustrated in Fig. 1. The dynamic antenna is based on a single printed meander line monopole antenna for near-omnidirectional radiation characteristics and miniaturization. We develop the structure into two tightly coupled meander line monopole antennas oriented in opposite directions with two-port excitation using image theory, which allows for a planar design on a single layer of substrate, suitable for simple front-end applications.

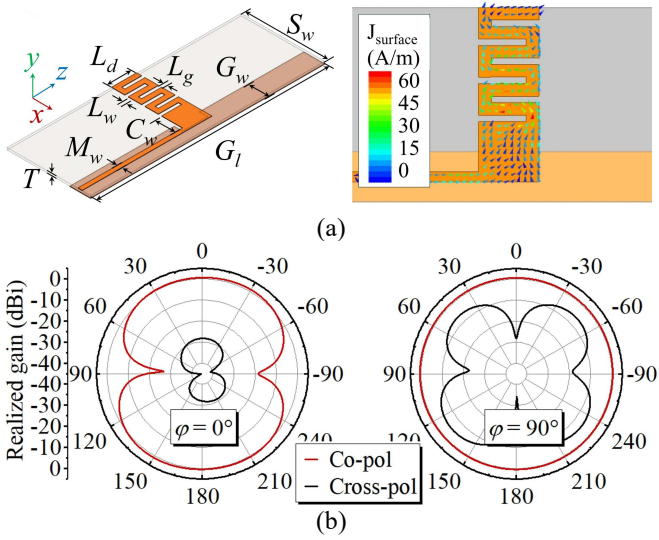

Fig. 2. The initial planar meander line antenna with microstrip feeding. (a) Top view and simulated current distribution at 3 GHz. (b) The corresponding co- and cross-polarization patterns. The design parameters are $L_g$ = 0.8, $L_w$ =1.2, $L_d$ = 6, $M_w$ =1.08, $G_w$ =5, $G_l$ =56, $T$ =0.508, $C_w$ = 5, and $S_w$ = 20 unit:mm.

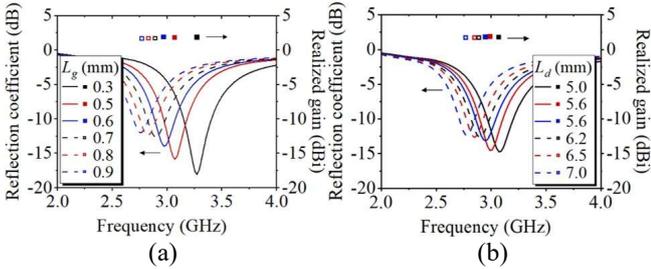

Fig. 3. The frequency tuning and the corresponding maximum realized gain of the single monopole antenna by (a) vertical length. (b) antenna width.

## II. THE DESIGN OF PLANAR MEANDER LINE ANTENNA

### A. Single Meander Line Monopole

The structure of the proposed planar meander line monopole antenna is shown in Fig. 2 (a). The low-cost fabrication and simple design process are achieved by printing the entire structure, including the meander line radiating element, capacitive patch, microstrip feeding and ground plane on a single layer of Rogers RO4350B substrate with dielectric constant of 3.48 and dissipation factor of 0.0037. A substrate thickness of $T$ = 0.508 mm is used. The design process of the meander line is well developed in [20], [34], where the meander structure is initially treated as two equivalent parts, a folded short-circuited terminal circuit, and a straight linear conductor. The antenna is fed by a 50 Ω microstrip line with line width of $M_w$. Then, a capacitive patch with length $C_w$ inspired by the work in [35] is used to miniaturize the antenna and to realize a practical feeding solution. The proposed meander line element has three meander sections and the line width and gap are indicated by $L_w$ and $L_g$. The capacitive patch and the meander line have the same overall width $L_d$. We optimized a long ground plane with length $G_l$ and width $G_w$ to ensure the maximum realized gain is at the broadside ($\theta$ = 0°). Fig. 2 (a) demonstrates the simulated surface current distribution of the proposed structure for the resonance at 3 GHz by using ANSYS HFSS, showing that it exhibits a typical current distribution of the meander line radiator. Tuning the resonance frequency of the antenna can be achieved by adjusting the horizontal width and the vertical length, which are mainly controlled by $L_d$ and $L_g$. Fig. 3 plots the simulated reflection coefficient and the maximum realized gain for varied $L_d$ and $L_g$. It is seen that for $L_g$ from 0.3 mm to 0.9 mm and $L_d$ from 5.0 mm to 7.0 mm, the resonance can be tuned from 2.75 GHz to 3.26 GHz, and the maximum realized gain can be maintained around 1.8 dBi.

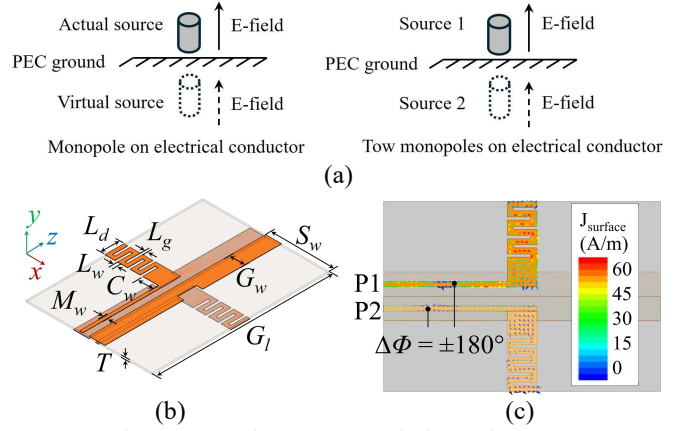

Fig. 4. The proposed structure of dynamic antennas. (a) Topology of two monopole antennas. (b) The proposed dynamic antenna on Rogers 4350B. (c) The simulated surface current at 3 GHz. Design parameters include $L_g$ = 0.65, $L_w$ = 1.25, $L_d$ = 6, $M_w$ = 1.08, $G_w$ = 5, $G_l$ = 56, T = 0.508, $C_w$ =5, and $S_w$ =20 unit:mm.

### B. Dual Monopole for Near-Omnidirectional Radiation

We expand the design into a dynamic antenna as shown in Fig. 4. Our work presented in [36] has shown that the spatial amplitude modulation of two antennas is an effective and simple technique for achieving a dynamic phase pattern. Here, we closely space two monopole antennas and leverage image theory for a vertical monopole antenna on an electric conductor [37]. To place the second actual source, an ideal 180° phase difference between two antennas should be considered so that the current on both conductors is in the same direction, acting as a single dipole antenna. In a practical design, the second antenna can be achieved by duplicating the initial single monopole, and therefore two antennas are realized in a single layer of substrate with radiating meander lines and ground planes mirrored to each other. Fig. 4 (c) shows the simulated surface current distribution at 3 GHz for two microstrip feeds P1 and P2 having $\Delta\phi$ = ±180° excitation, showing the current on both meander conductors are in the same direction.

We fabricated the antenna as shown in Fig. 5 (a) by a chemical etching process. An MHF4-RP SMA female jumper cable was used for feeding. The connection to the microstrip line was achieved by a soldered SMD Caps, compatible with the I-PEX MHF4 connector. The measured and simulated reflection coefficient are plotted in Fig. 5 (a), where a 50 Ω termination was loaded to either of the jumper cables. Both measured S11 and S22 have around 0.1 GHz shift mainly due to the parasitic inductance introduced by the cables and the soldered receptacle. The corresponding simulated and measured realized gain radiation pattern of the two states in the E- and H-planes are plot in Fig. 5 (b). In the measurement, we placed the adapters orthogonally to the electrical polarization to minimize the influence due to the reflection. For the E-plane ($\varphi = 0°$) in Fig. 5 (b), the measured beam shapes of State1 and State2 are symmetrical which match to the simulation. For the H-plane ($\varphi = 90°$) in Fig. 5 (b), the beam shapes of two states are almost static and the measured maximum realized gain is about 1 dBi lower than the simulation as them SMD Mount Receptable introduces 0.8 dB attenuation. To achieve a differential phase design by flipping the two port excitations, resulting in directional modulation, the radiation characteristics for differential amplitude feeding of Port 1 and Port 2 are important [36]. We investigate the relationship between the differential phase of State 1 and State 2 in two planes and the varied power ratio of two ports as shown in Fig. 5 (c). The E-plane differential phase slope is calculated based on the phase angle within $\theta = \pm 80°$, covering the beamwidth, and the H-plane differential phase slope considers phase at all angles. It can be seen that in the E-plane, differential phase occurs as long as the two ports are excited with unequal power. The minimum differential phase slope is obtained at -1.66 when $\alpha = \pm\infty$ dB. Meanwhile, the H-plane differential phase slope remains flat around 0 for all power ratios between the two ports, indicating static phase patterns when switching between the two states. The influence of the varied power ratio on maximum realized gain of two planes is also plotted, where the gain slightly drops by about 1 dB as the E-plane half power beamwidth increases as the power ratio increases. The E-plane maximum gain has a second peak value for $\alpha = \pm\infty$ dB as the main beam is split.

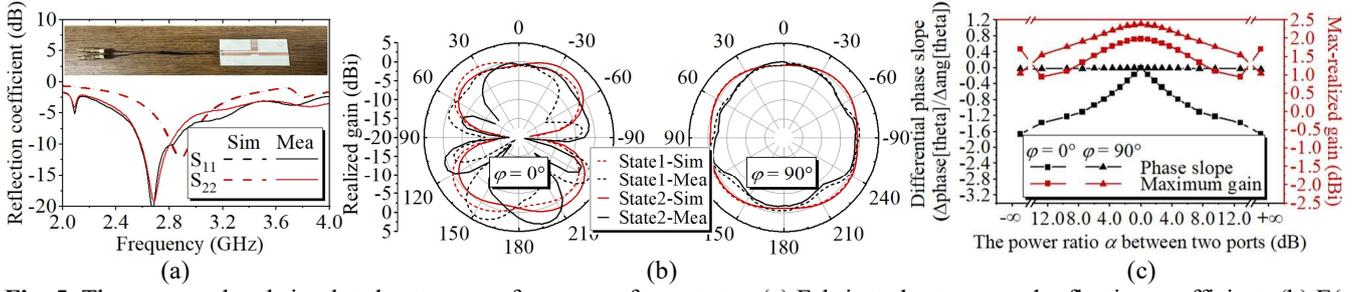

**Fig. 5.** The measured and simulated antenna performance of two states. (a) Fabricated antenna and reflection coefficient. (b) E($\varphi$ = 0°)- and H($\varphi$ = 90°)-plane realized gain patterns. (c) The influence of power ratio on phase slope and maximum gain.

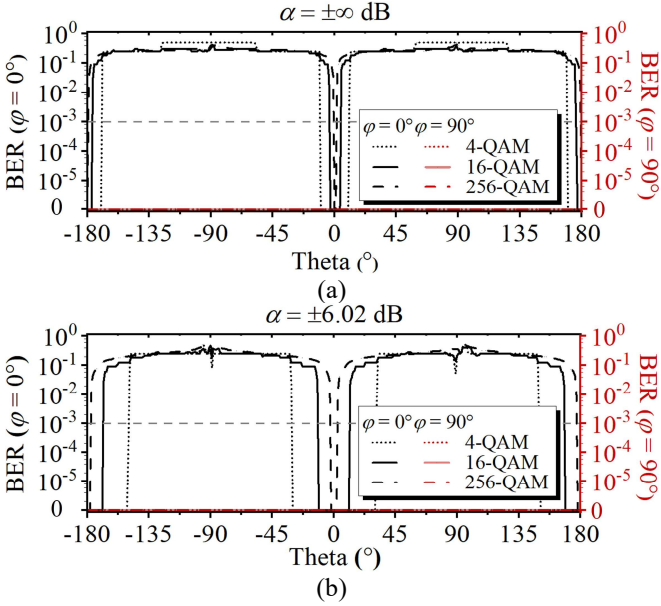

**Fig. 6.** The simulated communication performance using 4-, 16-, and 256-QAM for (a) $\alpha = \pm\infty$ dB and (b) $\alpha = \pm 6.02$ dB.

## III. EXPERIMENT IN WIRELESS COMMUNICATION

### A. The Investigation of Information Beam

We study the communication performance of the designed dynamic omnidirectional meander antenna using MATLAB based on the communication channel model in [36]. A 48 kb pseudorandom bit sequence was modulated onto a QAM signal using Gray coding and then the simulated amplitude and phase patterns of the two antenna states exported from HFSS were implemented. A SNR of 40 dB was set that ensures the all bit errors are due only to the directional modulation and not due to low SNR. We evaluate the information beamwidth (IB), defined as the angular separation for BER ≤ $10^{-3}$, to indicate the region where the transmitted information can be recovered. In Fig. 6, we plot the simulated BER in the E- and H-planes for 4-, 16-, and 256-QAM signals for amplitude ratios of $\alpha = \pm\infty$ dB and ±6.02 dB between two states. As expected, all simulation results confirm that a high BER above $10^{-3}$ is only observed in the E-plane while low BER (equal to 0) is obtained at all angles in the H-plane. This validates the concept that switching the feeding between two ports yields directional modulation in the E-plane only while the pattern in the H-plane remains static. It is apparent that high order QAM gives a narrower IB while for the same QAM order, a higher power ratio provides narrower IB results in a smaller recoverable region due to the lower differential phase slope.

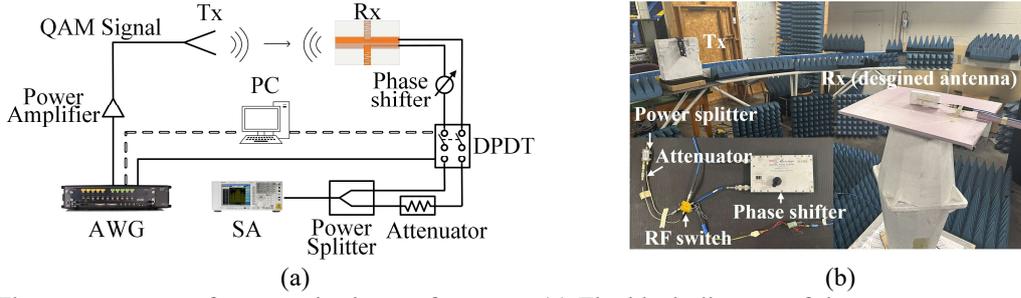

Fig. 7. The measurement of communication performance. (a) The block diagram of the measurement system configuration (b) The measurement setup.

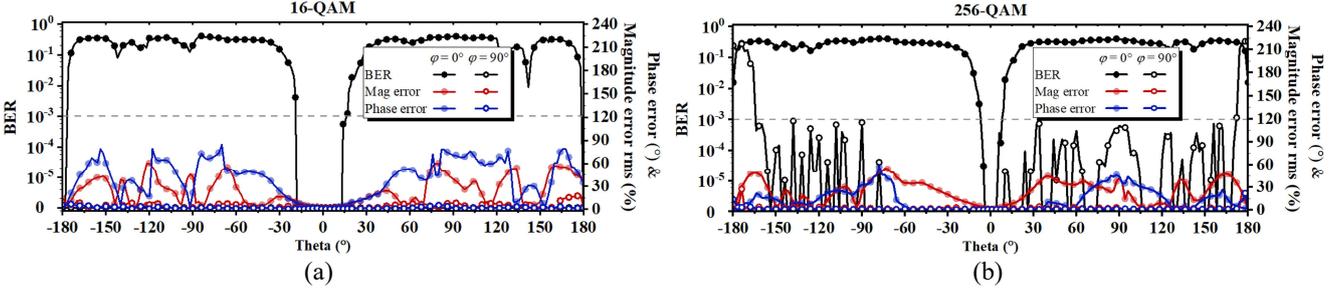

Fig. 8. The measured BER, magnitude error, and phase error of the omnidirectional antenna in E- and H-plane for (a) 16-QAM and (b) 256-QAM.

*B. Experiment of Communication Performance*

The block diagram and a photo of the measurement setup for measuring the BER of the dynamic antenna are shown in Figs. 7 (a) and (b). The dynamic antenna, along with the associated switching hardware, was on the receiving side. A communication signal at a 2.7 GHz carrier frequency was generated from a Keysight M8190A Arbitrary Waveform Generator (AWG) and amplified by a Mini-Circuits ZX6083LN-S+. The signal was transmitted via a commercial ultra wideband Vivaldi antenna (TSA800). A Narda Microline coaxial phase shifter was used to calibrate the relative phasing of the two feeds into the dynamic antenna. A total of 10 dB of differential amplitude between the feeds was achieved by 1 dB, 3 dB, and 6 dB attenuators applied to one path. The power ratios between the two switching states was calibrated at 9.1 dB and 10.0 dB; the asymmetry was due to differences in the path losses. The DPDT RF switch was used for switching the two RF paths with a 1 kHz switching frequency using a square wave from the AWG as a control signal. A power splitter (Mini-circuits ZX10-2-183-S+) combined the two RF receiving paths, after which the resultant signal was digitized. The signals were modulated and demodulated by Keysight IQtools and PathWave Vector Signal Analysis software.

The measured communication performance of the dynamic antenna using 16-QAM and 256-QAM signals at 2.7 GHz are shown in Figs. 8 (a) and (b), respectively. A Pseudo-Random Binary Sequence (PRBS) with period $2^{11} - 1$ and length of 30000 symbols was transmitted. The BER was measured in both the E-plane ($\varphi = 0°$) with 1° increment and the H-plane ($\varphi = 90°$) with 2° increments. At each angle, the BER was measured based on a total of 48k bits of data. Both measured BER results of 16-QAM and 256-QAM show that the designed dynamic antenna exhibits a narrow IB in the E-plane and effectively an omnidirectional IB in the H-plane. Specifically, the E-plane IB for 16-QAM was $-18° \leq \theta \leq 16°$ and $-176° \leq \theta \leq -180°$, and the E-plane IB for 256-QAM was $-7° \leq \theta \leq 8°$. The higher measured BER for θ around 180° occurred due to the connectors and feeding cables. On the other hand, an omnidirectional H-plane IB was obtained with BER at 0 for 16-QAM, while for 256-QAM, the IB was $-164° \leq \theta \leq 172°$. The corresponding measured magnitude and phase error at each angle are also plotted, which provide clear insight into the intrinsic dynamics of BER variation. For both 16-QAM and 256-QAM measurements, a high BER above $10^{-3}$ in the E-plane was due to the increased magnitude error and phase error, and a low BER under $10^{-3}$ was obtained where both magnitude error and phase error were close to 0. In the H-plane of the 16-QAM measurement, both magnitude error and phase error were close to 0 at all angles, resulting in an omnidirectional recoverable region, while for 256-QAM, the errors increase around $\theta = \pm 180°$ as the feeding cable is pointing to the transmitting antenna.

IV. CONCLUSION

In this work, a novel dynamic near-omnidirectional antenna integrated with directional modulation for secure planar wireless transmission has been demonstrated. By utilizing meander line monopole elements with dynamic port switching, the antenna achieves spatially selective information recovery in the E-plane while maintaining a highly flattened omnidirectional pattern in the H-plane. The antenna was fabricated and experimentally tested using 16-QAM and 256-QAM modulation schemes. The results demonstrate clear direction-dependent demodulation characteristics, where information recovery is confined to narrow angular sectors, validating the effectiveness of the proposed approach.